\documentclass[review, 11pt]{elsarticle}

\usepackage{graphicx}
\usepackage[usenames,dvipsnames]{color}
\usepackage[normalem]{ulem}

\usepackage{hyperref}

\newcommand{\Tix}[1]{(TiZrNbCu)$_{1-#1}$Ni$_#1$}
\newcommand{\Ti}{\Tix{x}}
\newcommand{\mOhm}{\,\mu\Omega\mathrm{cm}}
\newcommand{\Ang}{\mathrm{\AA}}

\newcommand{\mNg}{N_{\gamma}(E_F)}
\newcommand{\mNE}{N(E_F)}

\newcommand{\tauso}{\tau_{\mathrm{so}}}

\newcommand{\mTc}{T_c}
\newcommand{\mHc}{H_{c2}}
\newcommand{\mHcT}{\mHc(T)}
\newcommand{\mRH}{R_H}

\newcommand{\Ng}{$\mNg$}
\newcommand{\NE}{$\mNE$}
\newcommand{\Tc}{$\mTc$}
\newcommand{\Hc}{$\mHc$}
\newcommand{\HcT}{$\mHcT$}
\newcommand{\RH}{$\mRH$}

\newcommand{\myscale}{0.5}

\begin{document}

\begin{frontmatter}

\title{Change of electronic properties on transition from high-entropy to Ni-rich \Ti\ alloys}

\author[pmf]{Marko Kuve\v{z}di\'{c}}
\author[pmf]{Emil Tafra}
\author[pmf]{Mario Basleti\'{c}\corref{corrauth}}
\cortext[corrauth]{Corresponding author}
\ead{basletic@phy.hr}
\author[pmfos]{Ramir Risti\'{c}}
\author[ifs]{Petar Pervan}
\author[ifs]{Vesna Mik\v{s}i\'{c} Trontl}
\author[mex]{Ignacio A. Figueroa}
\author[pmf]{Emil Babi\'{c}}

\address[pmf]{Department of Physics, Faculty of Science, University of Zagreb, Bijeni\v{c}ka cesta 32, 10000 Zagreb, Croatia}
\address[pmfos]{Department of Physics, University of Osijek, Trg Ljudevita Gaja 6, HR-31000 Osijek, Croatia}
\address[ifs]{Institute of Physics, Bijeni\v{c}ka cesta 46, P. O. Box 304, HR-10001 Zagreb, Croatia}
\address[mex]{Institute for Materials Research-UNAM, Ciudad Universitaria Coyoacan, C.P. 04510 Mexico D.F., Mexico}

\begin{abstract}
We present results of comprehensive study of electronic properties of \Ti\ 
metallic glasses performed in broad composition range $x$ 
encompassing both, high entropy (HE) range, and conventional Ni-base alloy concentration range, $x \ge 0.35$. 
The electronic structure studied by photoemission spectroscopy and low temperature specific heat (LTSH) reveal a split-band structure of
density of states inside valence band with $d$-electrons 
of Ti, Zr, Nb and also Ni present at Fermi level \NE, 
whereas LTSH and magnetoresistivity results show that variation of \NE\ with $x$ 
changes in Ni-base regime. The variation of superconducting transition temperatures with $x$ 
closely follows that of \NE. 
The electrical resistivities of all alloys are high and decrease with increasing temperature over most of explored temperature range, and
their temperature dependence seems dominated by weak localization effects over a broad temperature range ($10 - 300\,$K). 
The preliminary study of Hall effect shows positive Hall coefficient that decreases rapidly in Ni-base alloys. 
\end{abstract}



\end{frontmatter}

\section{Introduction}
Recent concept of high entropy alloys (HEA, multicomponent near to equi-atomic alloys) \cite{1,2,3,4} has aroused large interest in the scientific community (see e.g., \cite{54}) 
The concept of HEAs is in contrast to that of conventional alloys, which are commonly based on a single primary element that has been doped in order to promote a particular desired property. 
This alloy design explores the middle section of the phase diagrams of multicomponent alloys which was largely unexplored in conventional alloys \cite{4}. Therefore, a virtually unlimited number of new alloys are now available for research and possible exploitation \cite{5}.

Intense research on HEAs over the last fifteen years has led to the production of several hundred new alloys \cite{54} and publication of about two thousand papers, including several reviews of literature (e.g., \cite{7,8,9,10,11,12,13,14,15}) and books \cite{16,17}.
As a result, large progress has been made in the research and understanding of HEAs and several technologically relevant alloys, such as those with exceptional low- and high-temperature mechanical properties have been discovered \cite{54}.

However, present research of HEAs seems somewhat uneven, with the main effort invested in the development of new structural materials, thus focused on their mechanical properties, microstructure and thermal stability (e.g., \cite{54,14,17}).
Meanwhile, the research aimed at their electronic structures and properties, both experimental \cite{54,18,19,20,21,22,23,24} and theoretical \cite{54,25,26,27,28,29}, is in spite of their potential as functional materials \cite{54} still insufficient.
The lack of insight into their electronic structure which in metallic systems determines almost all their properties (e.g., \cite{30a,30b}) hinders the conceptual understanding of both crystalline (c-) and amorphous (a-) HEAs \cite{19,24}.
Furthermore, the distribution of research on HEAs in regards to their composition is highly uneven, with studies of HEAs based on 3d-metals forming a large majority, while there are only a few studies on HEAs composed of early and late transition metals (e.g., \cite{1,2,54,17,19,22,24,31}).
In general, in spite of their early appearance \cite{1,2} and large conceptual importance\cite{17,24,31} research of a-HEAs  lags well behind that of c-HEAs.
This may be influenced by the fact that glass forming ability (GFA) of known a-HEAs is inferior in comparison to the best glass former in a particular alloy system \cite{15,17,32}.
However, in the last couple of years several comprehensive studies showing novel effects in a-HEAs have been published (e.g., \cite{32,33,34,35,36,37}).

The important problem of the transition from a-HEAs to conventional metallic glasses (MG) with the same chemical make-up was however ignored. The study of this transition is important both for understanding the formation of HEAs, and for proper evaluation of their potential in respect to that of conventional alloys.

Very recently, we reported, to our knowledge, the first systematic study of the atomic structure-electronic structure-property relationship in \Ti\ ($x \le 0.5$) MGs for $x$ in both the HEA ($x \le 0.35$) and conventional Ni-rich concentration range \cite{22,24}. 
It was found that both the atomic and the electronic structure show a change in the Ni-rich concentration range (valence electron number, $VEC \ge 7.4$ \cite{22}).
This was reflected in all properties studied, including so-called boson peaks, by the change of their concentration dependence for $x > 0.35$. 
The results have been compared with those for corresponding binary and ternary MGs \cite{24} sharing similar electronic structure.

Here we report the first, to the best of our knowledge, comprehensive study of the electronic properties of quinary MG system \Ti\ over a broad composition range ($x \le 0.5$) encompassing both the high-entropy and the conventional alloy concentration range ($x > 0.35$). Electrical resistivities of all alloys are high ($\rho \ge 160\mOhm$) and show qualitatively the same variations with $x$ and temperature as in corresponding binary MGs \cite{38,39a,39b,40,41,42}. Superconducting transition temperatures \Tc\ decrease with $x$ as expected from the split-band electronic density of states revealed by ultraviolet photoemission spectroscopy (UPS). However, in agreement with the change of atomic and electronic structure on transition from a-HEA to Ni-based MGs \cite{24} the decrease of \Tc\ with $x$ declines for $x > 0.35$. We also show that superconductivity in our alloys as well as that in crystalline HEAs \cite{18,21,23} and binary transition metal alloys (e.g.\ \cite{41}) is determined by their electronic structure and shows universal variation with the band (bare) density of states. The Hall coefficients, \RH, of all alloys are positive and decrease towards zero in Ni-rich alloys.

\section{Experimental}
As described in some detail elsewhere \cite{19,24}, eight alloys in the \Ti\ system with $x = 0$, 0.125, 0.15, 0.20, 0.25, 0.35, 0.5 and 0.55 were prepared from high purity elements by arc melting in a pure argon gas environment.
Thin ribbons with thickness of about $20\,\mu$m were fabricated from fragments of these alloys by melt spinning molten alloys on the surface of a copper roller in a pure He atmosphere \cite{19,22}.
Detailed description of the methods employed for structural (X-ray powder diffraction, XRD), chemical (scanning electron microscopy with energy dispersive spectroscopy, SEM/EDS) and thermal (differential scanning calorimetry and thermogravimetric analysis, DSC-TGA) characterisation of all alloys was reported previously \cite{19,22,24}.
XRD and DSC results revealed that all samples except for those with $x=0$ and 0.55 were fully amorphous. 
Therefore, the samples with $x = 0$ and 0.55 were not used in further studies. 
Details of the characterization of fully amorphous samples including their XRD patterns, SEM/EDS images and DSC curves were recently published \cite{19,22,24}. 
We note that all these samples showed homogeneous distribution of elements with actual compositions close to nominal ones. 
The techniques employed for the measurements of their low temperature specific heat (LTSH), magnetic and mechanical properties were also reported recently \cite{19,22}.

As described previously in \cite{24}, the valence-band structure of the as-cast sample was studied by UPS, with a Scienta SES100 hemispherical electron analyzer attached to an ultra-high vacuum chamber, with base pressure maintained below $10^{-9}\,$mbar. An unpolarized photon beam of $21.2\,$eV was generated by a He-discharge UV source. Several cycles of sputtering with $2\,$keV Ar$^+$ ions at room temperature were performed in order to remove oxygen and other contaminants from the sample surface. The energy resolution was about $25\,$meV.

The resistivity measurements were performed by the low frequency (typically 22$\,$Hz) ac method with rms current $i=0.2\,$mA. About $8\,$mm long samples for resistivity measurements were mounted on the sample holder of a $^3$He cryostat inserted into a 16/18 T Oxford superconducting magnet. The current and voltage wires were glued with silver paste onto the samples. The silver paste was allowed to dry at room temperature for about twelve hours resulting in contact resistances at most a few Ohms. The resistivity was determined from the measurements of resistance, length, mass and density of samples \cite{42,43,43b}. Due to finite width of the silver paste contacts the uncertainty in the absolute resistivity values was about 5\%. Additional uncertainty comes from voltage noise due to sizable Ag paste contact resistance. Compared to full voltage across the sample this noise was small (a few percents), i.e.\ less than the size of symbols in figures showing resistivities and superconducting transition. However, in a case of small signals such as the Hall voltage and magnetoresistance contributions (not shown) the noise was considerable part of the signal ($\simeq 10\%$) and is shown as the error bars. Raw UPS data are used so the error can be judged by the scatter in the intensity. As already stated the energy resolution of UPS was $25\,$meV, and is achieved by careful calibration of the setup.
This uncertainty propagated into the values of dressed density of states \NE\ determined from the resistivity and upper critical field \Hc\ results. The measurements were performed in the temperature range $0.3 - 300\,$K in magnetic field $B$ perpendicular to the broad surface of the ribbon and to the current direction. The temperature was measured with a calibrated Cernox thermometer situated close to the samples. The Hall effect was measured on the same samples used for resistivity measurements, but with contacts arranged in the Hall geometry, at several temperatures between $1.4\,$K and $300\,$K.

\section{Results and discussion}
As noted earlier \cite{19,22,24}, the values of thermophysical parameters such as the configurational entropy $\Delta S_{\mathrm{conf}}$, the mixing $\Delta H_{\mathrm{mix}}$ or formation enthalpy $\Delta H_{f}$ and the average difference in the atomic sizes of the constituents $\delta$, which are frequently employed in a semi-empirical criteria for the formation of different phases in HEAs \cite{13}, place \Ti\ within the range of intermetallic compounds forming alloys. This is probably associated with strong interatomic interactions $\Delta H_{\mathrm{mix}} < 0$ and large atomic size difference $\delta$ between the early (Ti,Zr,Nb) and late (Ni,Cu) transition metal components \cite{24}. In this sense, these alloys can be viewed as quasi-binary alloys of early (TE) and late (TL) transition metals which will facilitate their comparison with conventional (binary and multicomponent) TE-TL alloys. Since a large (negative) $\Delta H_{\mathrm{mix}}$ and $\delta$ also promote GFA in alloys \cite{13,19} the fabrication of amorphous alloys seems the simplest way to obtain \Ti\ alloys in a single phase state. Indeed, careful characterization of our samples \cite{19,22,24} revealed that alloys with $x = 0.125$, 0.15, 0.20, 0.25, 0.35 and 0.5 were fully amorphous and that the distribution of the constituents within the samples was random. In what follows, we discuss only the results for fully amorphous, as-cast samples.

\begin{figure}
\centerline{\includegraphics*[scale=\myscale]{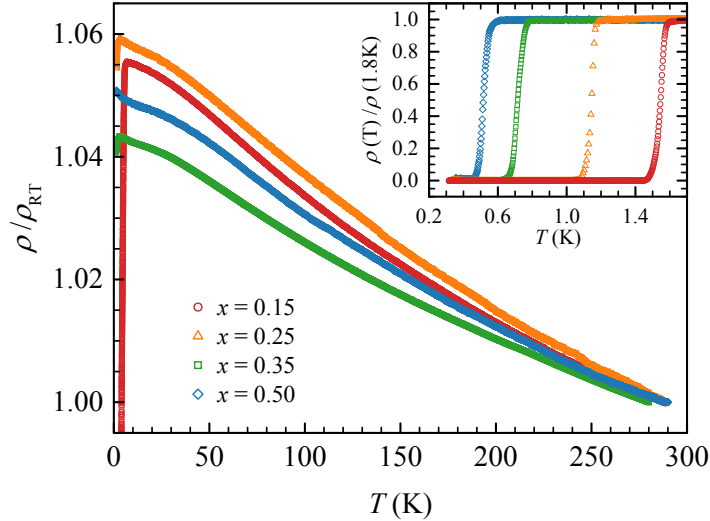}}
\caption{\label{F1} Normalized resistivities vs.\ temperature of selected \Ti\ amorphous alloys (Inset: superconducting resistivity transitions for the same alloys).}
\end{figure}

Figure \ref{F1} shows the variation of normalized resistivity with temperature for selected \Ti\ alloys. As is usual for both binary (e.g., \cite{38,39a,39b,40,41,42,43,43b,44,44b}) and multicomponent (e.g., \cite{45}) TE-TL amorphous alloys with high resistivity ($\rho \ge 140\mOhm$) the resistivities of all our alloys decreased with increasing temperature above $4\,$K.
This correlation between the magnitude of electrical resistivity and the corresponding temperature coefficient of resistivity, $\mathrm{TCR} = d(\ln R)/dT$, at room temperature in concentrated disordered transition metal alloys was first observed by J.\ H.\ Mooij \cite{50N}. He attributed the transition from $\mathrm{TCR} > 0$ to $\mathrm{TCR} < 0$ at elevated resistivities to a small electronic mean free path, $l$, in a way similar to that used in Ioffe-Reggel criterion \cite{51N}. (This criterion providing an estimate of the minimum metallic conductivity can be written in a form $k_F l \ge 1$, where $k_F$ is the Fermi wave-number which means that irrespective of the type of scattering, $l$ should not become smaller than the electronic wavelength. In other words, $k_F l < 1$ is criterion for an onset of electron localization.) 
Both, the values of the resistivities and their semiconductor-like variations with temperature are long-standing problem and several models have been proposed for their explanation (e.g., \cite{44,44b}). To the best of our knowledge the main mechanism(s) giving rise to the observed resistivities of amorphous TE-TL alloys has not been identified yet (e.g., \cite{42,44,44b}), whereas their temperature dependences seem to be dominated with the quantum interference effects (e.g.\ \cite{39a,39b,40,44}). The resistivities of our alloys, $166\mOhm \le \rho \le 176\mOhm$\ at room temperature, showed within a sizable scatter, $\Delta \rho \approx \pm 10\mOhm$, caused by a finite width of silver paste voltage contacts, a tendency to increase a little with $x$ from $x = 0.125$ to $x = 0.35$ ($\rho = 176\mOhm$) and then decreased at $x = 0.5$ ($\rho = 170\mOhm$). Such values of resistivities and their variation with $x$ are typical for binary TE$_{1-x}$TL$_x$ amorphous alloys \cite{38,39a,39b,40,44,44b} and were found to be insensitive on a type of TL for TL = Cu or Ni. Moreover, some other properties of TE$_{1-x}$TL$_x$ alloys such as the Debye temperature ($\Theta_D$), Young’s modulus and even the superconducting transition temperature $T_c$, are quite insensitive to the replacement of Ni with Cu. Because of this, in order to simplify the comparison of the results for our alloys with those for TE$_{1-x}$TL$_x$ alloys \cite{24} it is convenient to express the stoichiometry of our alloys in a form (TiZrNb)$_{(1-x_{\mathrm{TL}})}$(TL)$_{(x_{\mathrm{TL}})}$ with total TL content $x_{\mathrm{TL}} = x(\mathrm{Cu}+\mathrm{Ni})= x +(1-x)/4$, thus $0.344 \le x_{\mathrm{TL}} \le 0.625$ for studied alloys.

\begin{figure}
\centerline{\includegraphics*[scale=\myscale]{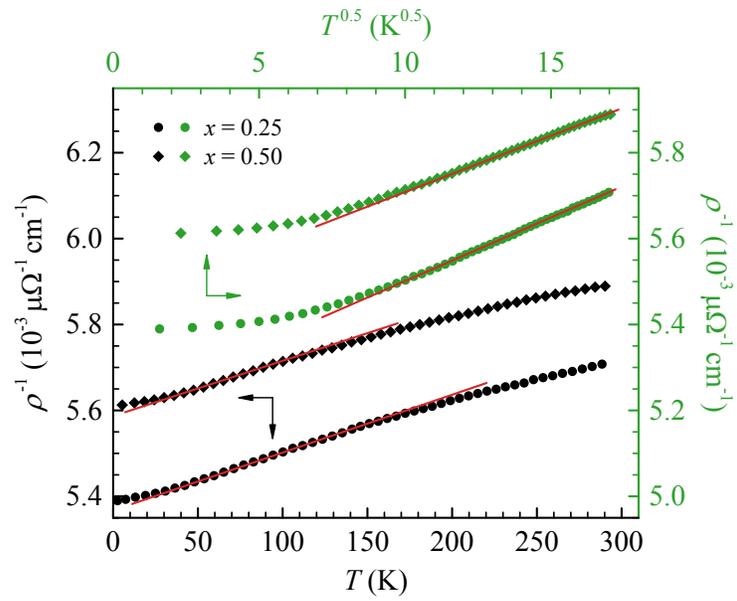}}
\caption{\label{F2} Left and bottom scale: $\rho^{-1}$ vs.\ $T$ for \Ti\ alloys with $x=0.25$ and 0.5. (Lines denote $\rho^{-1} \propto T$ dependence.) Top and right scale: $\rho^{-1}$ vs.\ $T^{0.5}$ for the same alloys. (Lines denote $\rho^{-1} \propto T^{1/2}$ dependence.)}
\end{figure}

Figure \ref{F2} shows the variations of electrical conductivity $\sigma = \rho^{-1}$ with temperature for the alloys from HEA ($x=0.25$) and Ni-rich ($x=0.5$) concentration range. $\sigma$ for both alloys shows asymptotic $T$ and $T^{0.5}$ (upper abscissa) variations for temperatures below and above about $100\,$K respectively. These variations of $\sigma$ probably reflect the temperature dependence of the inelastic scattering time of electrons in non-magnetic alloys \cite{39a,39b,40,44}: $\tau_i^{-1} \sim T^2$ at lower temperatures and $\tau_i^{-1} \sim T$ for $T > 100\,$K and follow from the expression for the electrical conductivity of three-dimensional disordered system in the regime of weak electron localization in the absence of magnetic field \cite{46}: 
\begin{equation}
	\Delta\sigma(H=0, T) = A\left( 2\sqrt{1+t} - \sqrt{t}\right)
\end{equation}
where $A = (\sqrt{3}e^2/2\pi l)\sqrt{\tau/\tau_{\mathrm{so}}}$, $t = \tau_{\mathrm{so}}/4\tau_i$, and $\tau$ and $\tau_{\mathrm{so}}$ are electronic transport life time and the life time due to spin orbit interaction, respectively.
The complete expression for $\sigma(T)$ in weakly localized regime \cite{46} includes in addition to delocalizing inelastic electron scattering $\tau_i$, and the effects of spin-orbit interaction $\tauso$, and also the effects of the electron-electron ($e-e$) interaction enhanced by disorder. We have shown earlier \cite{40} that this expression provides a good, quantitative fit to experimental $\sigma(T)$ results for amorphous Zr$_{42}$Cu$_{58}$ alloy over a full explored temperature range, $4.2 - 150\,$K. The effects of $\tauso$ and ($e-e$) interaction dominate at the lowest temperatures and cause the upward deviation of $\sigma(T)$ from linearity (lower part of figure \ref{F2} and \cite{40}). However, a self-consistent determination of both $\tau_i$ and $\tauso$ requires the accurate measurements of the magnetoresistance at low temperatures. Our preliminary measurements of the low temperature magnetoresistance (MR) of the sample with $x = 0.125$ showed qualitatively the behaviour which was predicted (Fig.\ 1 in \cite{46}) in case of weak localization and was experimentally observed (e.g.\ \cite{44, 53N}): rapid increase of MR at lower fields $\mu_0 H$, followed by saturation at elevated fields. Such variation of MR with $\mu_0 H$ is very different from simple $(\mu_0 H)^2$ variation arising from the classical Lorentz force contribution (e.g.\ \cite{54N}). Unfortunately, rather large scatter of our experimental MR data (probably caused by the sizable resistance of the silver paste contacts and the effects of elevated fields) prevented their detailed analysis. Still, qualitatively the same variation of $\sigma(T)$ and MR observed in our and other TE$_{1-x}$TL$_x$ amorphous alloys indicates that the effects of incipient localization dominate the temperature dependence of resistivity in all these alloys. Some arguments \cite{39a,39b,40} in support of this claim are: (1) very short electronic mean free paths, (2) a broad composition range showing a semiconducting resistivity behaviour, (3) direct relationship between the magnitude of increase of $\sigma$ with $T$ and electron-phonon interaction (figures \ref{F1} and \ref{F2} and \cite{39a,39b,40}), (4) large low-temperature magnetoresistance which cannot be accounted for with classical Lorentz force contribution, etc. All these arguments apparently apply to our alloys.

As illustrated in the inset to Figure \ref{F1} all samples were superconducting with transition temperatures $T_c \le 1.6\,$K. The resistive superconducting transitions were sharp, $\Delta T_c \le 0.07\,$K, where $\Delta T_c$ is the temperature interval between 0.1 and $0.9R(1.8\,$K). These sharp transitions, much sharper than those observed in Zr-based bulk metallic glasses \cite{45}, probably confirm good quality of our samples in accord with the results of XRD and SEM/EDS studies \cite{19,22}. As already noted \cite{19}, such low $T_c$s, comparable to these in amorphous Hf$_{1-x}$Cu$_x$ alloys \cite{41}, are unusual considering rather high contents of Nb (pure Nb has $T_c = 9.2\,$K), Zr and Ti which form alloys with $T_c$s around $10\,$K, but are consistent with the results for amorphous Nb-Zr films which showed no superconductivity down to $T = 1\,$K. The monotonic decrease of $T_c$ with $x$ in our and all other TE$_{1-x}$TL$_x$ amorphous alloys (e.g., \cite{41,43,43b}) can be simply explained in terms of their electronic structures.

\begin{figure}
\centerline{\includegraphics*[scale=\myscale]{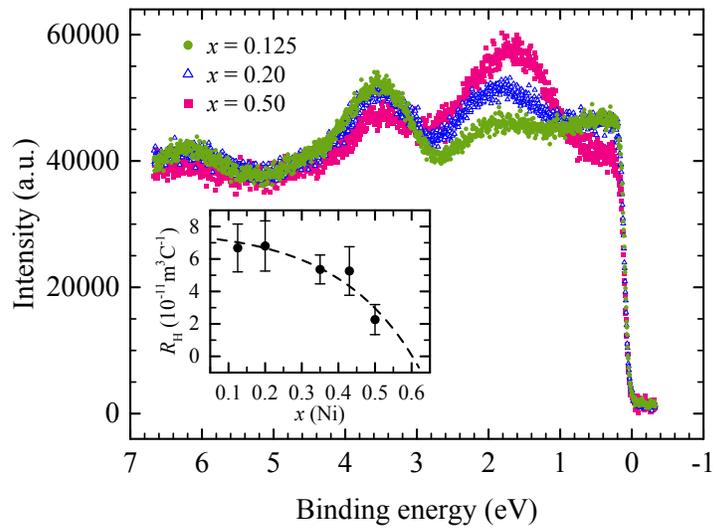}}
\caption{\label{F3} UPS spectra for \Ti\ samples with $x=0.125$, 0.2 (data from \cite{24}) and 0.5. Inset: Hall coefficients $R_H$ for \Ti\ alloys vs.\ Ni concentration $x$. (Dashed line is a guide to the eye.)}
\end{figure}

\begin{figure}
\centerline{\includegraphics*[scale=\myscale]{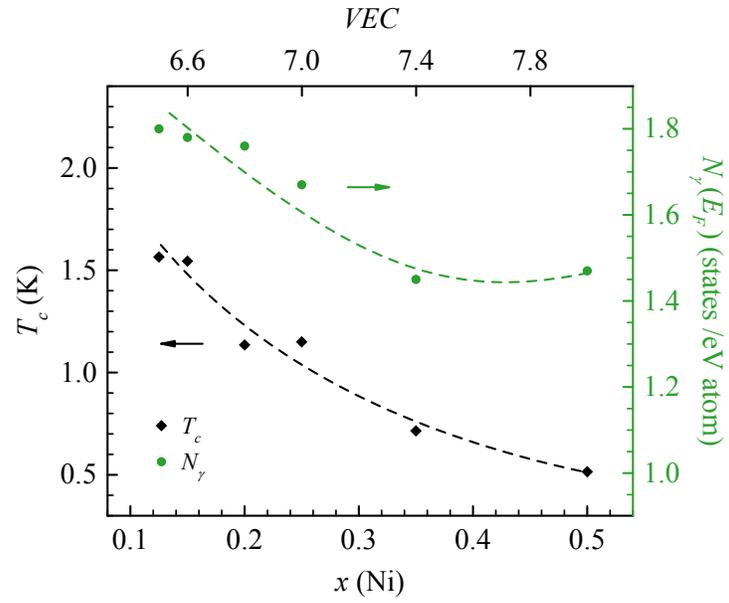}}
\caption{\label{F4} Left scale: $T_c$ vs.\ $x$, right scale: dressed density of states vs.\ $x$ for all \Ti\ alloys (see text for details). Upper abscissa shows $VEC$. (Both dashed lines are a guides to the eye.)}
\end{figure}

Figure \ref{F3} shows the UPS spectra for \Ti\ alloys with $x=0.125$, 0.2 and 0.5. These spectra reflect the evolution of the electronic density of states (DOS) within the valence band on increasing Ni content. All three spectra show clearly split-band structure of valence band with Cu-3$d$ states and Ni-3$d$ states giving rise to spectral maxima at $3.5\,$eV and $1.7\,$eV below the Fermi level ($E_F$). The spectral intensities at and close to $E_F$ are dominated with $d$-states of TEs; Ti, Zr and probably to some lower extent of Nb \cite{24}. The Ni contribution to the intensity at $E_F$ is not negligible however and probably increases with increasing $x$. However, spectrum for $x=0.5$ ($x_{\mathrm{TL}}=0.625$) does not indicate band-crossing which was observed in Zr$_{1-x}$Ni$_x$ alloys with $x \ge 0.64$ \cite{24,47,47b,48}. In particular, $E_F$ of this alloy seems to remain in the area dominated with $d$-states of TEs. A more quantitative insight into the evolution of DOS at $E_F$ of our alloys with $x$ has been obtained recently from the LTSH measurements \cite{22}. These measurements have shown that the coefficient $\gamma$ of the electronic contribution to LTSH, which is proportional to dressed DOS at $E_F$, \Ng\ ($\gamma = \pi^2 k_B^2 \mNg/3$), where $k_B$ is the Boltzmann constant and $\mNg = N_0(E_F)(1+\lambda_{e-p})$ where $\lambda_{e-p}$ represents the electron-phonon enhancement of DOS at $E_F$, $N_0(E_F)$, changes its variation with $x$ for $x > 0.35$. In particular, $\gamma$ \cite{22} and also \Ng\ (figure \ref{F4}) decrease approximately linearly with $x$ for $x \le 0.35$ (as expected from their split-band structure of DOS, figure \ref{F3}) but seem to stop changing at $x=0.5$. The values of $\gamma$ for our alloys when plotted vs.\ $x_{\mathrm{TL}}$ nearly overlap those for binary amorphous Zr-Cu, Ni alloys \cite{24} for $x_{\mathrm{TL}} \le 0.513$ ($x=0.35$), but deviate strongly upwards from binary alloy data for $x_{\mathrm{TL}} = 0.625$ ($x=0.5$). We note that the change in variation of $\gamma$ and \Ng\ with $x$ coincides with that in structural parameters of the same alloys occurring for $x_{\mathrm{TL}} > 0.513$ (which corresponds to $VEC \ge 7.4$) \cite{22,24}.

As seen in figure \ref{F4} the variation of $T_c$s (defined as the midpoint of the resistive transitions in zero applied field, $0.51\, \mathrm{K} \le T_c \le 1.57\,$K) of our alloys with $x$ and $VEC$ (upper abscissa) is qualitatively the same as that of \Ng. (Probably better definition of $T_c$, free from the effects of a flux flow and superconducting fluctuations, would be as a temperature at which a deviation from the normal state behaviour appears, but in our measurements the scatter in the experimental resistivity data prevents an accurate determination of this temperature.) This indicates that superconductivity in \Ti\ amorphous alloys irrespective of their composition and structural change obey the Dynes-Varma correlation \cite{49} which seems valid for all disordered transition metal alloys, both crystalline (e.g., \cite{21,23}) and amorphous \cite{41}. As already noted in \cite{24}, since \Ng\ is enhanced by the electron-phonon interaction, it cannot be used to prove that the changes in figure \ref{F4} for $x > 0.35$ are due to change in $N_0(E_F)$. However, in superconducting transition metal alloys one can use the McMillan expression \cite{50} in order to disentangle $N_0(E_F)$ from $\lambda_{e-p}$ in \Ng. By using the data from figure \ref{F4} we calculated the values of $N_0(E_F)$ for all our alloys which confirmed a change in $N_0(E_F)$ for $x>0.35$. We also found that both, the values of $N_0(E_F)$ (ranging from 1.19$\;$states/eV$\,$atom for $x=0.125$ to 1.01$\;$states/eV$\,$atom for $x=0.5$) and  $\lambda_{e-p}$ (ranging from 0.51 for $x=0.125$ to 0.41 for $x=0.5$) of our alloys are somewhat lower than those for corresponding Zr-Cu, Ni amorphous alloys (e.g.\ \cite{42}) which, together with the adverse influence of Ti \cite{41,49}, probably explain their low $T_c$s. Simultaneously, the initial rates of decrease of \Tc\ and \Ng\ with $x_{\mathrm{TL}}$ and $VEC$ of our alloys are very close to those in Zr$_{1-x}$Ni$_x$ alloys \cite{43,43b} which implies that the same mechanism is responsible for the initial suppression of \Tc\ in both alloy systems.

\begin{figure}
\centerline{\includegraphics*[scale=\myscale]{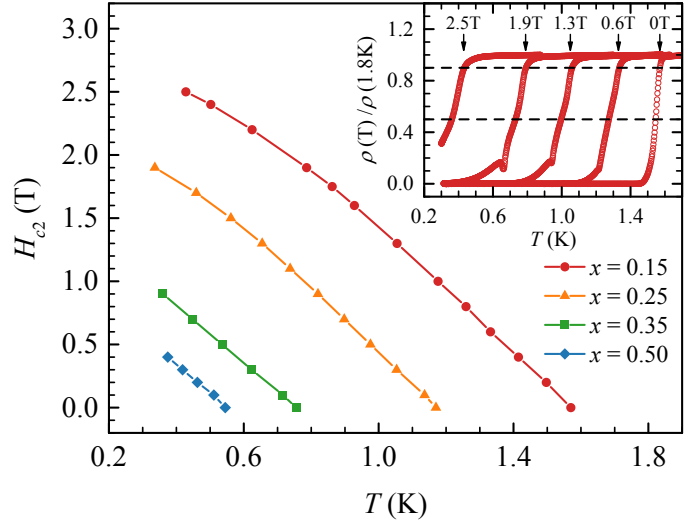}}
\caption{\label{F5} Upper critical fields \HcT\ vs.\ $T$ for selected \Ti\ alloys. (Lines are a guides to the eye.)
Inset: superconducting transition in alloy with $x = 0.15$, for different fields.}
\end{figure}

Useful information about superconductivity in amorphous alloys can be obtained from the measurements of the upper critical field \HcT\ (e.g.\ \cite{41}). In particular, in the absence of LTSH results one can estimate \Ng\ from the measurements of \HcT\ and $\rho$:
\begin{equation}\label{eq1}
\mNg = - \frac{\pi}{4 e k_B N_A} \frac{M}{\rho d} \left(\frac{\mathrm{d}\mHc}{\mathrm{d}T}\right)_{T = T_c} \ ,
\end{equation}
where $N_A$, $M$ and $d$, are the Avogadro number, molecular weight and mass density respectively, and other symbols have their usual meaning \cite{41}. The \HcT\ variations for selected \Ti\ alloys encompassing both HEA and Ni-base concentration range are shown in figure \ref{F5}, \Hc\ was defined with resistivity criterion $\rho(\mHc) = 0.9 \rho(1.8\,\mathrm{K})$, but as illustrated in the inset, due to sharp resistive transitions some other definitions of \Hc\, such as $0.5 \rho(1.8\,\mathrm{K})$ would cause very small downward shift of the data in figure \ref{F5}. We note that magnetic field shifts and broadens the resistive transition but the broadening, caused by a weak flux pinning inherent to homogeneous amorphous superconductors, is the strongest for $\rho < 0.5 \rho(1.8\,\mathrm{K})$ therefore its effect on the \HcT\ variation is negligible.

The initial slopes of the upper critical field $(\mathrm{d}\mHc/\mathrm{d}T)_{T_c}$ had values typical for amorphous TE$_{1-x}$TL$_x$ alloys ($2.3-2.6\,$T/K) and decreased with $x$, but this decrease stopped for $x > 0.35$. By using (\ref{eq1}) we have calculated \Ng\ from the corresponding results for $(\mathrm{d}\mHc/\mathrm{d}T)_{T_c}$  and $\rho$. In spite of the rather large scatter, caused by the uncertainty in $\rho$, the calculated values of \Ng\ are close to those obtained from LTSH measurements (figure \ref{F4}). In particular, these values of \Ng\ also showed a tendency to saturation for $x \ge 0.35$.

As is well known, \Hc\ also provides important information on the nature of superconductivity in the studied system \cite{41}. In particular, \Hc\ provides the information on coherence length $\xi$, the fundamental length scale for superconducting state:
\begin{equation}\label{eq2}
\xi(T) = \sqrt{\frac{\Phi_0}{2\pi B_{c2}(T)}}
\end{equation}
where $\Phi_0 = h/2e$ is the magnetic flux quantum. Further, \Hc\ provides important information on the mechanisms causing the destruction of superconductivity in applied magnetic field. However, due to low \Tc s, the data for \HcT\ of our samples does not cover a sufficiently broad range of reduced temperatures $t = T/T_c$ which is required in order to obtain reliable results to all parameters affecting \HcT\ variation \cite{41}. Because of this we focused on obtaining reliable estimates of $\mHc(0)$ which provides an insight into the mechanism leading to the destruction of superconductivity. Accordingly, we performed fits of $\mHc(t)$ data for samples with $x \le 0.25$ for which both the explored $t$ range and the number of data points seem adequate (figure \ref{F5}) for a reliable fitting. Since the fits of our data to the phenomenological expression $\mHc(t) = \mHc(0)(1-t^2)$ were not adequate we used a more general expression  $\mHc(t) = \mHc(0) + Bt + Ct^2$ which provided good fit to experimental results. The obtained $\mHc(0)$ values were close to Pauli limited field $\mHc(0) = B_{\mathrm{Pauli}} \approx 1.83 T_c$ where 1.83 comes from the BCS gap ratio, which indicates that depairing occurs when Zeman energy splitting becomes comparable to the energy gap. Thus, providing that $\mHc(t)$ for Ni-based alloys also extrapolate close to $B_{\mathrm{Pauli}}$ for $t \rightarrow 0$, the $\mHc(0)$ values follow the same variation with $x$ as \Tc\ (figure \ref{F4}). The values of $\mHc(0)$ ranged from $2.9\,$T for $x=0.125$ to $0.94\,$T for $x=0.5$, with corresponding coherence length $107\,\Ang$ and $187\,\Ang$, respectively, see Eq.\ (\ref{eq2}).

\begin{figure}
\centerline{\includegraphics*[scale=\myscale]{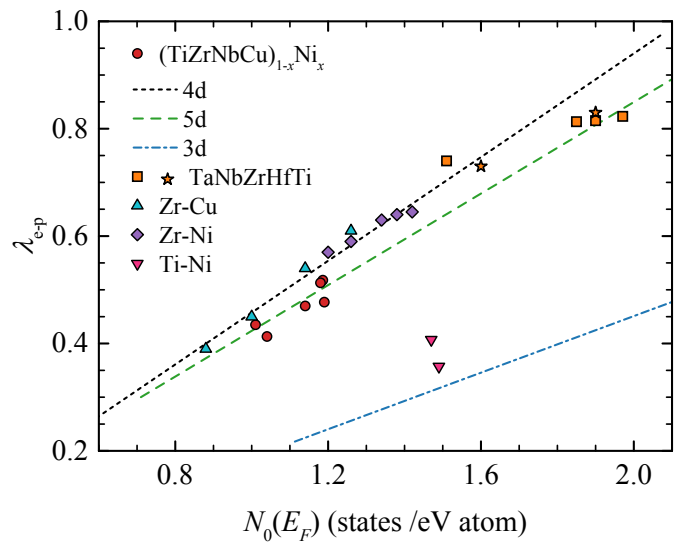}}
\caption{\label{F6} Electron-phonon coupling constant vs.\ band density of states at the Fermy level for disordered transition metals and their alloys. Circles represent our data; lines are from ref. \cite{49}; squares are from \cite{18,21}; stars are from \cite{23}; up-triangles are from \cite{51}; diamonds are from \cite{43,52}; down-triangles are from \cite{53}. (Dashed lines corresponds to prediction in Ref.\ \cite{49}.)}
\end{figure}

As already noted and shown explicitly in figure \ref{F6}, our superconducting results obey quite well the Dynes-Varma correlations \cite{49} for disordered transition metal alloys. According to this correlation the superconductivity in such alloys is determined with $N_0(E_F)$, which result in a linear variation of $\lambda_{e-p}$ with $N_0(E_F)$ with a slope that depends only on a class of transition metals (i.e., 3$d$, 4$d$ or 5$d$) and the position of $E_F$ within a given $d$-band density of states \cite{49}. As seen in figure \ref{F6}, within the experimental scatter, our results accomodate quite well with the previous results for 5$d$-alloys \cite{49}, in spite of the fact that they do not contain 5$d$-metals. This probably arises from significant amount of Ti (3$d$-metal), which suppresses strongly the rate of increase of  $\lambda_{e-p}$ with $N_0(E_F)$ as illustrated with the literature results for amorphous Ti-Ni alloys \cite{48} in figure \ref{F6}. Similarly, the results for crystalline (bcc) Ta-Nb-Zr-Hf-Ti HEAs (which contain components from all transition metal series) are extending between the lines drawn previously for 4$d$ and 5$d$-alloys \cite{49} and their position depends on the fraction of 3$d$, 4$d$ and 5$d$ metal. In particular, alloy with lowest Ti content is closest to 4$d$-line, whereas these with highest Ta content are located at the 5$d$-line. We note that results for HEAs in figure \ref{F6} extend the validity of Dynes-Varma correlation to mixtures of different transition metals.

The Hall coefficient of \Ti\ alloys studied so far is positive and has the values similar to those observed in amorphous Zr-Co, Ni, Cu alloys, $R_H \le 6.8\times 10^{-11}\,\mathrm{m}^3\mathrm{C}^{-1}$ \cite{47,47b}. As seen in the inset to Fig.\ \ref{F3}, $R_H$ is practically constant for $x \le 0.2$ and decreases rapidly for $x \ge 0.43$, reaching $2.2\times 10^{-11}\,\mathrm{m}^3\mathrm{C}^{-1}$ for $x = 0.5$. A similar decrease of $R_H$ in amorphous Zr-Ni, Co alloys has been associated with approach to crossing of $E_F$ from Zr $d$-band to the solute $d$-band around a certain concentration $x_c$. Linear extrapolation of the variation of $R_H$ between $x=0.35$ and $x=0.5$ to $R_H = 0$ yielded $x_c = 0.61$ which is consistent with the approximate expression for $x_c$ developed in \cite{47,47b}. Thus, the variation of $R_H$ in multicomponent TE-TL amorphous alloys is, as in corresponding binary alloys \cite{47,47b}, determined by the changes in their electronic structure. However, the origin of $R_H > 0$ in these alloys is still not completely understood \cite{42}.

\section{Conclusion}
Main results of the comprehensive study of electronic properties of \Ti\ amorphous alloys over a broad composition range (encompassing both high-entropy and conventional alloy concentration range $x>0.35$) are presented and compared with those for corresponding binary and multicomponent amorphous alloys of early (TE) and late (TL) transition metals.

The observed magnitudes of the electrical resistivities ($\rho$) as well as their variations with $x$ and temperature ($T$) are very similar to those in corresponding binary TE$_{1-x}$TL$_x$ alloys, which probably implies that chemical complexity has little effect on electronic transport in amorphous TE-TL alloys. In particular, both in binary and our quinary alloys the variations of $\rho$ with $T$ seem dominated with the quantum interference effects which show up in the magnetoresistance and specific variations of $\rho^{-1}$ vs.\ $T$ at lower and higher ($T \ge 100\,$K) temperatures, respectively. The observed superconducting transition temperatures \Tc\ are surprisingly low ($T_c \le 1.57\,$K) and decrease further with increasing $x$, as expected from their split-band electronic density of states with $d$-states of TEs dominant at the Fermi level ($E_F$), as demonstrated by the photoemission spectra for $x=0.125$, 0.2 and 0.5. The decrease of \Tc\ with total TL content $x_{\mathrm{TL}}$ and $VEC$ within the high-entropy concentration range were practically the same as in binary Zr$_{1-x}$Ni$_x$ alloys implying the same mechanism for the suppression of superconductivity. However, in agreement with the change in atomic and electronic structure on transition from HEA to Ni-based alloys, the decrease of \Tc\ with $x$ declined for $x>0.35$. Thus, \Tc\ is determined by the density of states at $E_F$ as is usual in disordered transition metal alloys. Moreover, both our and the literature results for crystalline HEAs show that superconductivity in alloys containing mixture of different transition metals is also determined with the band density of states. Since the estimated upper critical fields at $T=0\,$K ($\mHc(0)$) seem to be Pauli limiting fields, the variation of $\mHc(0)$ with $x$ is probably the same as that of \Tc. The preliminary results for the Hall effect show that $R_H>0$ and has a similar magnitude as that in Zr$_{1-x}$TL$_x$ alloys with low $x$. However, $R_H$ shows a tendency to change a sign to $R_H<0$ for $x>0.5$, which probably indicates that the relationship between the $R_H>0$ and band-crossing is the same as in binary Zr$_{1-x}$TL$_x$ amorphous alloys.

Summarizing the above, we note that the electronic properties of all disordered alloys of early and late transition metals strongly depend on their electronic structure and quenched-in disorder, and are less sensitive on their chemical complexity, i.e.\ number of components. (Indeed, recent study of the atomic structure, electronic structure and magnetism in similar (TiZrNbNi)$_{1-x}$Cu$_x$ MGs \cite{63N} revealed an ideal solution behaviour similar to the observed in binary TE-Cu MGs \cite{42}.) Therefore, the change in their properties on transition from HE to conventional alloy concentration range depends primarily on the corresponding change in their electronic and atomic structure.

\section*{Acknowledgments}
Our research was financially supported by the Josip Juraj Strossmayer University of Osijek, and by the University of Zagreb through funds for multipurpose institutional financing of scientific research. We also acknowledge the support of project CeNIKS co-financed by the Croatian Government and the European Union through the European Regional Development Fund - Competitiveness and Cohesion Operational Programme (Grant No. KK.01.1.1.02.0013). I. A. Figueroa acknowledges the financial support of UNAM-DGAPA-PAPIIT.

\bibliography{biblio}

\end{document}